\documentclass[fleqn,10pt]{wlscirep}

\usepackage{arydshln}
\usepackage{graphicx}
\usepackage{color}
\usepackage{booktabs}
\usepackage{tikz}
\usepackage{multirow}
\usepackage{rotating}
\usepackage{enumitem}

\newcommand{\mc}{\multicolumn}
\newcommand{\mcc}{\multicolumn{2}{c}}
\newcommand{\mccc}{\multicolumn{3}{c}}

\newcommand{\tb}{\textbf}

\newif\iffigsinpdf

\title{QuickProbs 2: Towards rapid construction of high-quality alignments of large protein families} 

\author[1,*]{Adam Gudy\'s}
\author[1]{Sebastian Deorowicz}

\affil[1]{Institute of Informatics, Silesian University of Technology, Akademicka 16, 44-100 Gliwice, Poland}
\affil[*]{adam.gudys@polsl.pl}


\begin{abstract}
Increasing size of sequence databases caused by the development of high throughput sequencing, poses multiple alignment algorithms to face one of the greatest challenges yet. As we show, well-established techniques employed for increasing alignment quality, i.e., refinement and consistency, are ineffective when large protein families are of interest. We present QuickProbs~2, an algorithm for multiple sequence alignment. Based on probabilistic models, equipped with novel column-oriented refinement and selective consistency, it offers outstanding accuracy. When analysing hundreds of sequences, QuickProbs~2 is significantly better than Clustal$\Omega$, the previous leader for processing numerous protein families. In the case of smaller sets, for which consistency-based methods are the best performing, QuickProbs~2 is also superior to the competitors. Due to computational scalability of selective consistency and utilisation of massively parallel architectures, presented algorithm is comparable to Clustal$\Omega$ in terms of execution time, and orders of magnitude faster than full consistency approaches, like MSAProbs or PicXAA. All these make QuickProbs~2 a useful tool for aligning families ranging from few, to hundreds of proteins.

QuickProbs~2 is available at \href{https://github.com/refresh-bio/QuickProbs}{https://github.com/refresh-bio/QuickProbs}.
\end{abstract}

\begin{document}
\flushbottom
\maketitle
\thispagestyle{empty}

\section*{Introduction}

Multiple sequence alignment (MSA) is of crucial importance in life sciences. The ability to reveal evolutionary and structural relationships between sequences makes MSA the basic tool in a number of biological analyses, including phylogeny, structure prediction, gene finding, and many others. Rapid dissemination of high throughput sequencing technologies causes sequence databases to grow exponentially~\cite{Kahn2011}. To face this, the development of alignment algorithms able to process thousands of sequences in a reasonable time is required.

Among many proposed heuristics for finding multiple sequence alignments, progressive scheme has become the most popular. It consists of three steps: (I) estimating evolutionary distances between sequences, (II) building a guide tree based on the distances, (III) greedy alignment of sequences in the order described by the tree. The classic representative of progressive aligners with more than 50 thousand citations (Google Scholar, June 2016) is ClustalW~\cite{Tho1994}. The greatest disadvantage of progressive algorithms is the propagation of mistakes from bottom levels of the guide tree to the final result. A lot of techniques were introduced to counter this issue. Historically, the first approach was to fix errors made at the progressive stage by iteratively refining output alignment~\cite{Gotoh1996}. This idea has been successfully acquired by a number of algorithms like MAFFT~\cite{Kat2002}, MUSCLE~\cite{Edgar2004}, or MSAProbs~\cite{Liu2010}. A different iteration scheme has been introduced in Clustal$\Omega$~\cite{Sievers2011} which combines recalculations of a guide tree and profile-HMM on the basis of a preliminary alignment. This results in superior accuracy for large protein families. An alternative way of facilitating progressive heuristics is to prevent mistakes during alignment construction. This can be achieved in various ways. One method is to employ information from suboptimal alignments, e.g., by calculating posterior probabilities on the basis of pair-hidden Markov models (ProbCons~\cite{Do2005}), partition function (ProbAlign~\cite{Ros2006}), or both of those (MSAProbs). Another is incorporating knowledge from other pairwise alignments when processing given pairs of sequences/profiles. The technique is known as consistency, and has originally been used in T-Coffee~\cite{Notredame2000}. Consistency has been proven to significantly elevate alignment quality and has been successfully acquired in different variants by a number of progressive (MAFFT~\cite{Kat2005}, ProbCons, MSAProbs), and non-progressive algorithms (PicXAA~\cite{Sahraeian2010}). However, a substantial drawback of consistency-based methods, i.e., excessive computational complexity with respect to the number of sequences, limits their applicability to families of approximately hundred of members. Consequently, algorithms allowing thousands or more sequences to be aligned like Kalign2~\cite{Las2009}, Kalign-LCS~\cite{Deo2013}, Clustal$\Omega$, or MAFFT in PartTree mode~\cite{Kat2007} do not use consistency. 


In the article, we give a new insight into the effect of refinement and consistency on progressive alignment. We investigated large sequence sets showing that accuracy of the aforementioned techniques scales unsatisfactorily with the number of sequences. In particular, when sets of hundreds or thousands of sequences are of interest, existing refinement variants have little effect on alignment quality, while consistency decreases it by introducing more noise than relevant information. We present new ideas to overcome those issues. i.e., column-oriented refinement and selective consistency. 

The research was based on QuickProbs algorithm~\cite{Gud2014} which is a successor of MSAProbs--one of the most accurate multiple sequence aligners~\cite{Liu2010}. Thanks to the utilisation of massively parallel architectures, QuickProbs is about order of magnitude faster than MSAProbs preserving quality of the results. We introduce QuickProbs~2, which is a significant improvement over its predecessor. Column-oriented refinement converges to alignments of higher quality than existing methods, while selective consistency incorporates most relevant information from pairwise alignments effectively reducing number of mistakes in a progressive scheme also for large sets of sequences. Moreover, selectivity decreases dramatically computational effort needed for consistency. This, together with optimised implementation, allows QuickProbs~2 to produce alignments superior to its forerunner at a fraction of a time. As a result, presented algorithm is the most accurate aligner when investigating protein families ranging from few to hundreds of sequences. Facilities like nucleotide mode or bulk processing further extend QuickProbs 2 usability.

\section*{Methods}

In the paper we introduce QuickProbs~2, a novel algorithm for multiple sequence alignment. It consists of four stages: 
(I) calculation of posterior probability matrices, 
(II) construction of the guide tree, 
(III) consistency transformation, 
(IV) construction of the final alignment followed by the iterative refinement.
Posterior probability matrices are calculated for all sequence pairs on the basis of hidden Markov model~\cite{Dur1998} and partition function~\cite{Miyazawa1995}. The matrices are further employed to establish maximum expected accuracy alignments. Alignment scores are used to estimate pairwise distances which are given as an input for weighted UPGMA algorithm~\cite{Sne1973} for guide tree construction. In order to incorporate information from all pairwise alignments when aligning given pairs of sequences/profiles, posterior matrices are relaxed by other sequences during consistency transformation. Then, the proteins are progressively aligned in the guide tree order with a use of relaxed posterior matrices. This is followed by the iterative refinement.   

The most important advances with respect to existing methods were achieved at stages III and IV. QuickProbs~2 has been equipped with a novel column-oriented refinement and selective consistency, which are described elaborately in following subsections. A separate subsection concerns other algorithmic improvements and new facilities introduced in the presented algorithm. Finally, we describe in detail benchmark datasets and measures used for quality assessment.

\subsection*{Column-oriented refinement}
Refinement was designed to overcome the most important disadvantage of progressive algorithms--misalignments caused by the propagation of errors from early progressive steps up the guide tree. 
Usually, the procedure employs an iterative scheme of alternate splits and realignments and incorporates an objective function for results evaluation. A number of refinement strategies were investigated in the literature~\cite{Hirosawa1995,Wallace2005,Chakrabarti2006b}. Amongst them random and tree-guided approaches have become the most common in MSA algorithms.

First revision of QuickProbs, similarly to ProbCons or MSAProbs, employs the former idea: each refinement iteration splits alignment randomly into two horizontal profiles and realigns them after removing columns containing only gaps. No objective function is incorporated. The substantial drawback of the procedure is that
the larger the number of sequences, the smaller the chance of producing profiles with gap-only columns. As a result, no columns are removed in the majority of cases and the realigned profile is likely to be the same as the input one. Therefore, for numerous sets consecutive random refinements give no improvement in accuracy. 
An alternative approach, incorporated e.g. by MUSCLE or MAFFT, is tree-guided refinement. It splits alignment by breaking randomly selected branch in the guide tree. As gap-only columns are more likely to occur due to gathering phylogenetically related sequences in subprofiles, this approach can potentially be more successful when large protein families are of interest.

In the research we present a new approach to refinement which considers columns containing at least one gap. The algorithm selects randomly one of those columns and splits alignment into two profiles depending on the gap presence in this column. 

\begin{figure}[thp]
	\centering	
	\includegraphics[width=0.66\textwidth]{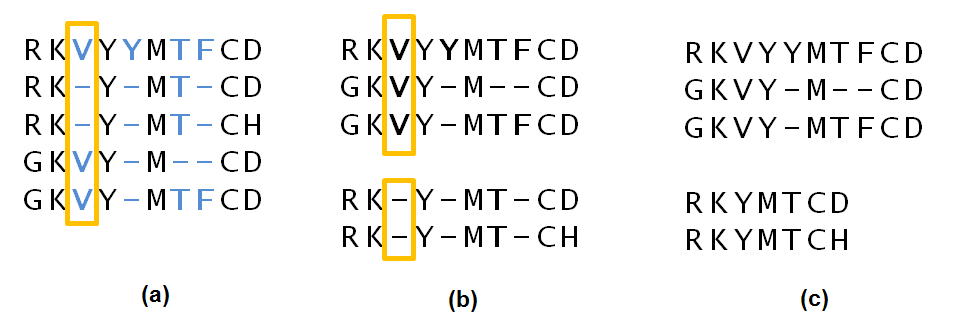}	
	\caption{Iteration of column-oriented refinement: (a) one of the candidate columns (in blue) is randomly selected as a splitter (bounded with an orange box); (b) the alignment is divided into two profiles according to the presence of gaps in the selected column; (c) gap-only columns are removed which is followed by profile realignment.}
	\label{Figure:refinement_example}
\end{figure}

As a result, at each refinement iteration at least one profile is shortened increasing significantly the chance of rearranging alignment and producing higher quality outcome. This type of refinement will be referred to as \emph{column-oriented} refinement and, as experiments show, it is superior to the random and tree-guided approaches, especially for large sequence sets. Figure~\ref{Figure:refinement_example} presents the application of column-oriented refinement on example alignment.

An important observation is that the number of gaps $g$ in a column, according to which an alignment is divided, affects sizes of resulting profiles. The closer is $g$ to the half of the sequence set size $k$, the more balanced is the division. 
To investigate the effect of imbalance in profile splitting on alignment quality
columns were sorted with respect to $|g - \frac{k}{2}|$. Then, only assumed fraction from the beginning or from the end was considered in the random selection (these correspond to the bias towards respectively, more or less balanced splits).

The refinement is often facilitated by introducing the objective function.
The usage of \textit{unsupervised} SP score under assumed substitution matrix and gap penalty model (not to confuse with \textit{supervised} SP score calculated on the basis of the reference alignment) is amongst the most popular~\cite{Edgar2004, Kat2005}.
Nevertheless, maximising unsupervised SP score does not necessarily converge to biologically meaningful alignments~\cite{Edgar2010,Wallace2005},
particularly for consistency algorithms~\cite{Wallace2005}.  
In the research, we suggest alignment length to be used as a straightforward and effective measure for refinement supervision. Intuitively, misalignments at consecutive progressive steps accumulate, causing blocks of conserved symbols to be shifted with respect to each other. As a result, one can expect erroneous alignments to be longer than those correctly identifying evolution of sequences. This hypothesis is supported by the observation that average alignment length declines as quality increases in consecutive refinement iterations. Therefore, we introduced to refinement an acceptance criterion of non-increasing alignment length which further increased the convergence.

In the research, we also examined entropy-based acceptance rule of non-decreasing \emph{trident score}~\cite{Valdar2002}. The method employs three components for column scoring: amino-acid conservation, stereochemical properties, and the presence of gaps.

\subsection*{Selective consistency}  
As opposed to refinement, consistency aims at preventing misalignments introduced in progressive scheme rather than eliminating them afterwards. To reduce the chance of making errors, consistency employs information from all pairwise alignments when aligning a pair of two particular sequences. Even though this approach has been successfully applied in a number of progressive MSA algorithms, the excessive computational cost limits its applicability to sets of approximately hundred of sequences. To our knowledge, the effect of consistency on larger sequence sets has not been investigated in the literature. \cite{Sievers2013} made an exhaustive study on scalability of MSA algorithms examining the effect of addition of homologous sequences to the reference set on the alignment accuracy. For all observed methods quality deteriorated when more than 50 sequences were added. The decay was especially steep for several consistency-based methods (e.g., MSAProbs, ProbCons) suggesting that for larger sets of sequences, noise exceeds relevant information. This, however, has not been explicitly verified.

In QuickProbs~2, similarly to its predecessor, consistency relies on the relaxation of posterior probability matrices by other sequences. The computational complexity of $\Theta(k^3n^3)$, with $n$ being the sequence length, makes this stage very time consuming\footnote{Precisely, posterior matrices are represented in a sparse form with a sparsity coefficient $\beta < 1$. As presented in supporting information to~\cite{Gud2014}, the time complexity depends on the structure of sparse matrices and varies from $\Theta(\beta^2 k^3n^3)$ to $\Theta(\beta k^3n^3)$}. Nevertheless, as QuickProbs comes with a fast relaxation algorithm suited for graphics processors, we were able to investigate the effect of consistency on sets exceeding thousand of sequences.
As presented in the experimental section, the procedure decreased alignment quality for protein families of such sizes.

\begin{figure}[t]
	\centering	
	\includegraphics[width=0.5\textwidth]{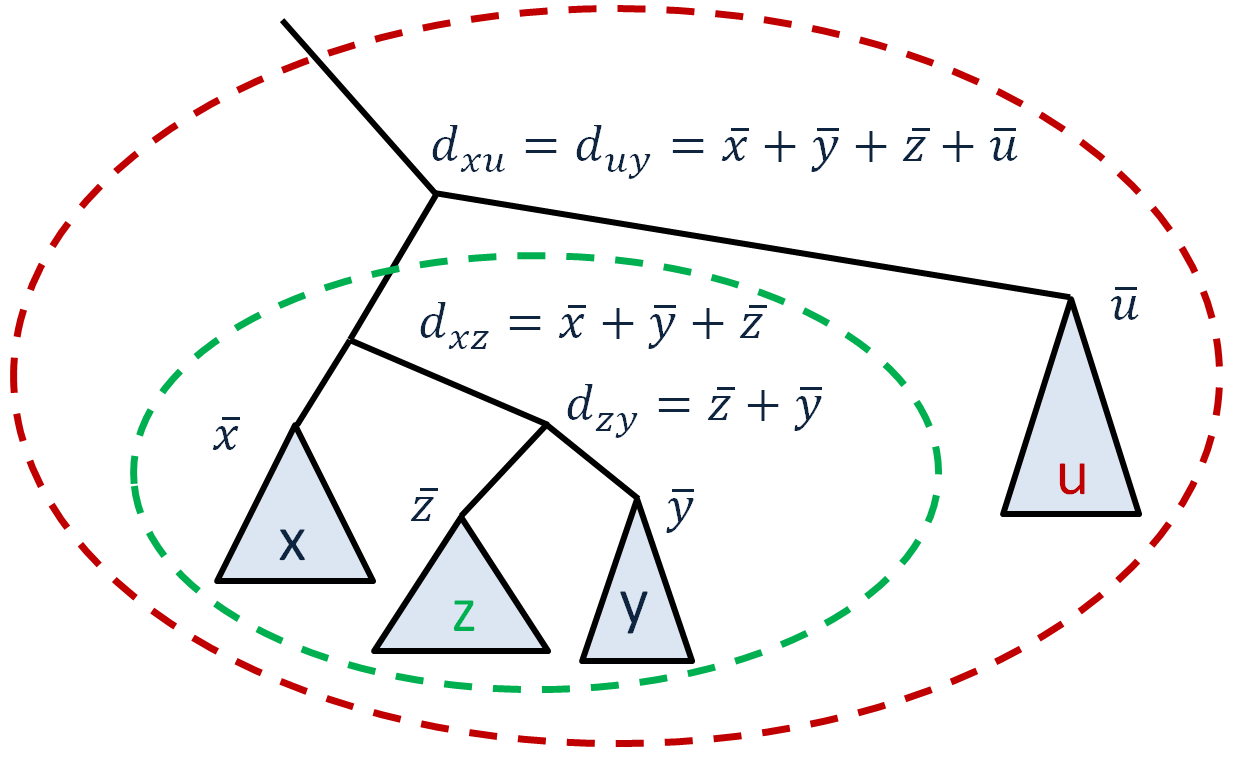}	
	\caption{Tree-guided selective consistency 
		of $S_{xy}$ posterior matrix with threshold $T$. Triangles represent subtrees with sequences $x$, $y$, $z$, and $u$. At each node, a size of a subtree is given. In the example, selectivity procedure accepts relaxation of $S_{xy}$ through $z$ as $f(d_{xz}, d_{zy}) \leq T$ (green oval). At the same time it excludes sequence $u$ from consistency due to $f(d_{xu}, d_{uy}) > T$ (red oval).}
	\label{Figure:consistency_example}
\end{figure}

The challenge which naturally arises, is to apply consistency only on sequences carrying most of the information. Particularly, we examined whether there is a correlation between information content and evolutionary relationship of sequences involved in consistency. For this purpose we introduce \textit{selective consistency}. Given $x$, $y$, $z$ sequences and $d_{xz}$ and $d_{yz}$ distances, posterior matrix $S_{xy}$ is relaxed over sequence $z$ if assumed function $f(d_{xz}, d_{yz})$ fulfills given condition. In the research we investigated two different $d_{xy}$ measures: (a) score-based distance calculated at stage I, ranked and normalised to $[ 0,1]$ interval, (b) tree-guided distance defined as a number of nodes in a minimal subtree containing both $x$ and~$y$. 
Maximum, minimum, or sum can be used as examples of $f$ function.
Selectivity was applied either by deterministically thresholding $f$ on arbitrary value or by applying stochastic filtering.
The \emph{filter function} maps a value of $f(d_{xz}, d_{yz})$ to the probability of performing consistency over sequence $z$. Its shape determines which sequences are preferred in the consistency procedure (e.g. closely or distantly related).
As thresholding tree-guided distances rendered superior results, we explain this variant of selectivity in Figure~\ref{Figure:consistency_example}.

The side effect of selectivity is the variability in the number of relaxations performed for different posterior matrices. Consequently, the larger the number of sequences undergoing consistency transformation, the weaker is the signal from original $S_{xy}$ compared to matrices it is multiplied by. 
To overcome this, we additionally analysed the effect of amplification of this signal on alignment quality by multiplying $S_{xy}$ elements by coefficient $h_{xy}$. The value of $h_{xy}$ varies from 1 when no relaxations of $S_{xy}$ are performed, to $h$ when maximum number of relaxations under chosen selectivity settings is done (200 in our case). This allows sets of different sizes to be handled properly.

\subsection*{Other algorithmic improvements}
In spite of focusing QuickProbs~2 research on extending refinement and consistency stages, calculation of posterior matrices was also a subject to some modifications. Quality improvements include replacing Gonnet160 matrix for partition function calculation by VTML200, which was proven to be more accurate~\cite{Muller2002}. This was followed by training partition function parameters, i.e., gap penalties and temperature on BAliBASE 3~\cite{Thompson1999} benchmark with a use of NOMAD algorithm~\cite{LeDigabel2011} for optimisation of non-smooth functions. Another changes were introduced in order to shorten execution time. They include redesigning graphics processor calculations to handle sequences of any length, optimisation of both CPU and GPU codes, and using more efficient memory allocation model. As a result, posterior calculation stage in QuickProbs~2 is more accurate than its predecessor, being at the same time several times faster. QuickProbs~2 is also equipped with a nucleotide mode in which HOXD substitution matrix~\cite{Schwartz2000} and GTR evolutionary model~\cite{Tavare1996} are used. Accurate mode, which in QuickProbs adjusted sparsity coefficient in posterior matrices, is no longer supported due to excessive computation time and lack of significant influence on the results.

Due to different behaviour of consistency depending on the set size, the number of transformations is adjusted to the number of sequences (2 for $k < 50$, 1 otherwise). 
It was also discovered that for two consistency transformations, 30 iterations of refinement instead of default 200 is sufficient to get satisfactory convergence.

\begin{figure*}[thp]
	\includegraphics[width=\textwidth]{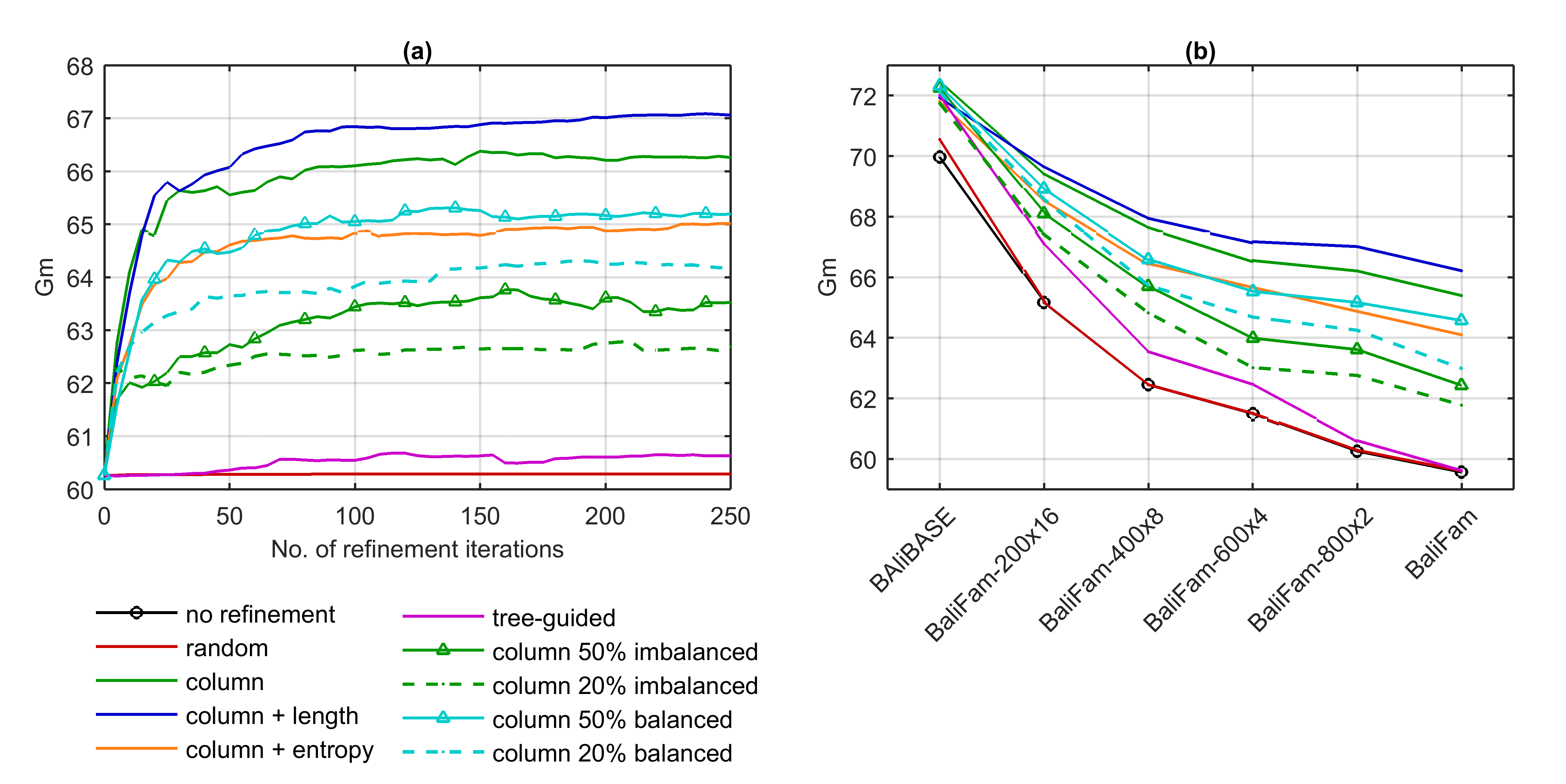}	
	\caption{Comparison of refinement strategies: (a) effect of consecutive iterations on BaliFam-800x2, (b) scalability with respect to the number of sequences in a set after 200 refinements.}
	\label{Figure:refinement}
\end{figure*}

As QuickProbs~2 employs OpenCL, it can be executed on different massively parallel devices like NVidia and AMD GPUs. Moreover, presented software has also the ability to be run on central processor without OpenCL. For convenience, QuickProbs~2 is also equipped with \emph{bulk mode} allowing any number of sequence sets to be processed during a single run. Necessity of storing posterior matrices for all pairs of sequences causes memory to be the major limiting factor for the set size. For this reason, QuickProbs~2 gives the opportunity to fit analysis in user-specified amount of RAM by decreasing sparsity coefficient in posterior matrices. Naturally, the adjustment affects quality and is possible only within certain boundaries.   

\subsection*{Accuracy assessment}

Accuracy of algorithms was assessed on several benchmark datasets that come with reference alignments. Those were BAliBASE~\cite{Thompson1999}, PREFAB~\cite{Edgar2004}, extended version of OXBench~\cite{Raghava2003}, SABmark~\cite{Walle2005}, HomFam~\cite{Blackshields2010}, and BaliFam~\cite{Sievers2013}. The four former were downloaded in a standardised FASTA format from Robert Edgar's Webpage~\cite{EdgarBench} and consist of small and moderate sequence sets (up to tens of sequences in the majority of cases). The latter were constructed by enriching respectively, Homstrad~\cite{Mizuguchi1998} and BAliBASE benchmarks, with full protein families from Pfam~\cite{Punta2012}. Number of sequences in BaliFam sets is in the order of 1,000 while Homfam contains much larger families of even 100,000 members. Both benchmarks were postprocessed by removing duplicated sequences which appear numerously due to generation protocol. 
 This was motivated by the fact that duplicates may negatively affect accuracy of analysed algorithms and can be straightforwardly restored after alignment has finished. 

Postprocessed BaliFam contained 218 sets with 934 sequences on average. 
As the major part the research focuses on the scalability of presented methods with respect to the number of sequences, 
BaliFam was recursively resampled to obtain less numerous sets: initial benchmark into two sets of 800 sequences, each of those into two sets of 600, and so on. Finally, elements at the same level of the pyramid were gathered forming sets referred to as BaliFam-800x2, BaliFam-600x4, BaliFam-400x8, and BaliFam-200x16. Such protocol has the property of smaller sets being contained in larger ones and preserves representativity for all problem sizes. 
As for the HomFam, afetr duplicate removal, all its sets were randomly downsampled to 1300 members with a guarantee of preserving sequences present in the reference alignments. 
This was motivated by the fact that original HomFam sets were too large to be processed by QuickProbs 2 due to memory requirements. Sampled benchmark will be referred to as HomFam$^\text{1K}$ and contained 94 families with 1093 sequences on average. Detailed histograms of familiy sizes in BaliFam and HomFam$^\text{1K}$ are presented in Supplementary~Figure~1.  

Quality evaluation was performed with well established metrics related to reference alignments. Those are supervised sum of pairs (SP) and total column (TC) scores defined as a fraction of correctly aligned symbol pairs and columns, respectively. When single quality measure was needed, e.g., for visualisation, geometric mean (Gm) of aforementioned scores was employed. Separate charts for SP and TC measures are given as Supplementary Figures.

\section*{Results}

\subsection*{Refinement}
In the initial experiments, we investigated random and tree-guided refinements together with different variants of novel column-oriented procedure. As refinement was acquired by alignment algorithms formerly to consistency, the latter was disabled in this experimental part. BaliFam-800x2 benchmark was selected as a representative of large protein families instead of BaliFam because it contains twice as many sets which reduces results variability. The effect of consecutive refinement iterations is presented in Figure~\ref{Figure:refinement}a, while scalability of refinement with respect to the set size after 200 iterations can be observed in Figure~\ref{Figure:refinement}b.

As charts show, for numerous protein families such as those in BaliFam-800x2, consecutive random refinements gave no improvement in accuracy. Moreover, random procedure was profitable only for BAliBASE and starting from BaliFam-200x16 it had no effect on the results. 
The performance of tree-guided refinement was noticeably better, however it also declined with the increasing number of sequences.
The opposite situation was in the case of column-oriented refinement. Not only it was superior to the competing approaches, but was also characterised by perfect scalability. Namely, its effectiveness raised from 2\% on BAliBASE to almost 7\% on BaliFam, confirming the selection of gap-only columns to be the choice for large protein families. 
When analysing the effect of split imbalance on alignment quality, it is visible that the bias towards more or less balanced columns (50\% and 20\% variants) caused accuracy decay. Consequently, the version without preference was chosen for further investigation. 

The final refinement experiments concerned the effect of different acceptance rules. Those were non-increasing alignment length and non-decreasing entropy score. As presented in Figure~\ref{Figure:refinement}, the former improved refinement convergence for larger sequence sets being only slightly inferior to unsupervised variant on BAliBASE. In contrast, entropy scoring performed unsatisfactorily on all analysed sets. As a result, column-oriented refinement with length supervision the was selected for QuickProbs~2. Charts presenting influence of refinement on SP and TC measures separately can be found in Supplementary Figure~3.

\subsection*{Consistency}

\begin{figure*}[t]
	\centering	
	\includegraphics[width=\textwidth,natwidth=162bp,natheight=227bp]{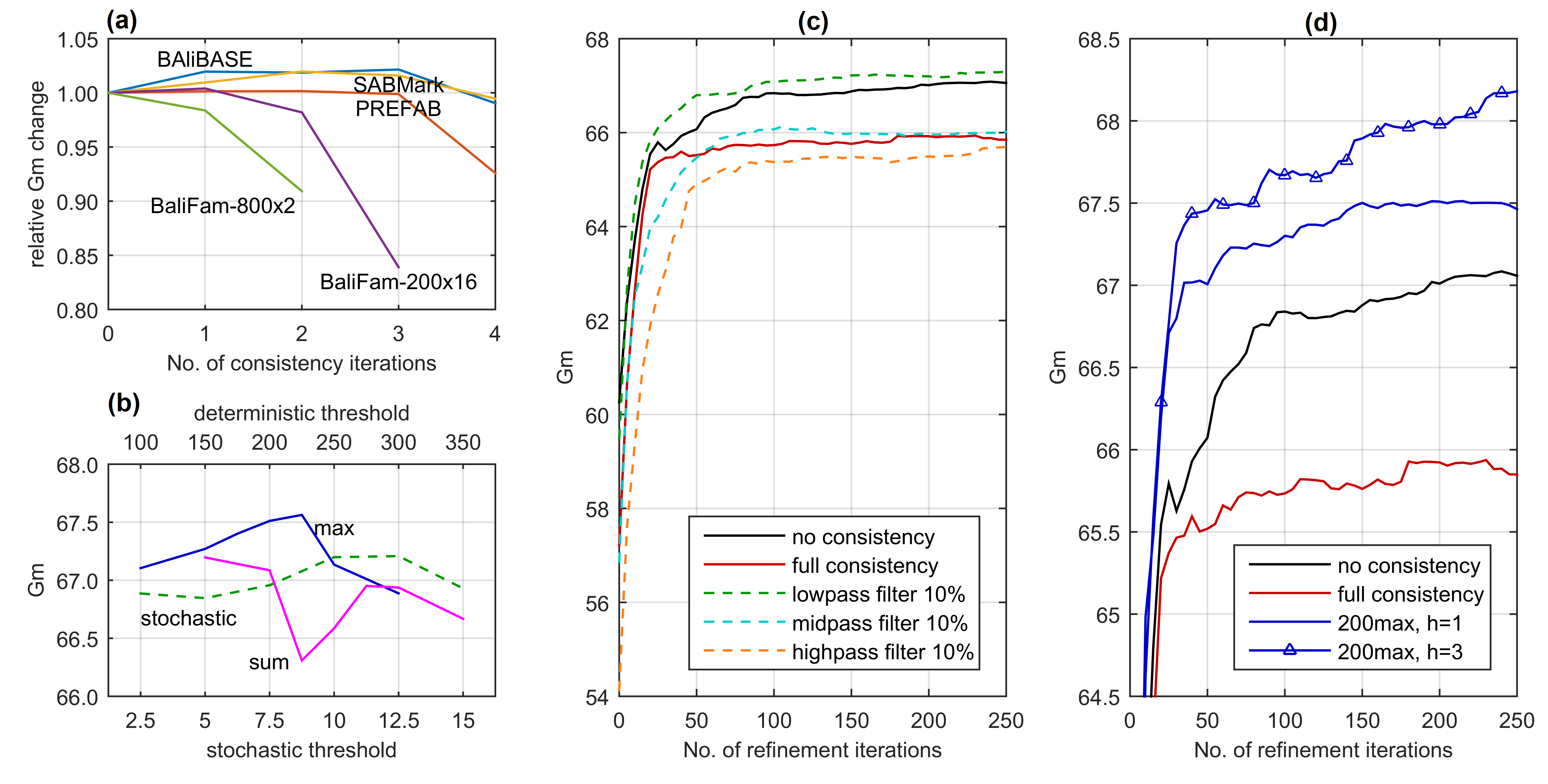}	
	\caption{(a) Effect of consistency iterations on selected benchmarks. Analysis of consistency on BaliFam-800x2: (c) effect of distance-related filtering, (b) selectivity variants for closely related sequences, (d) weighting original posterior matrices by $h_{xy} \in [ 1, h ]$ coefficient. Alignment qualities from charts (a) and (b) measured after 200 refinements.}
	\label{Figure:consistency_details}
\end{figure*}

\begin{figure*}[t]
	\centering	
	\includegraphics[width=\textwidth,natwidth=162bp,natheight=227bp]{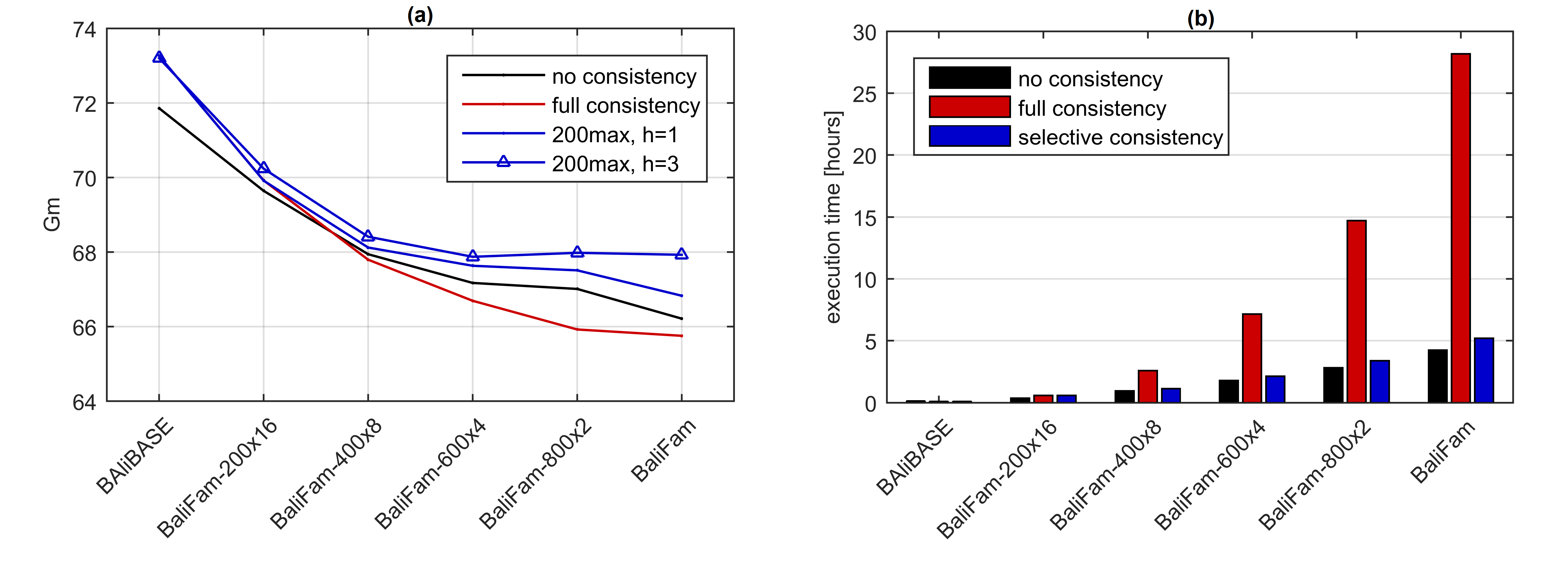}	
	\caption{Scalability of consistency after 200 refinement iterations: (a) alignment quality, (b) execution time on \textit{desktop} configuration with GeForce GPU.}
	\label{Figure:consistency_scalability}
\end{figure*}

Another experimental part concerned the analysis of consistency. Figure~\ref{Figure:consistency_details}a shows the effect of traditional (non-selective) consistency iterations on selected benchmarks after 200 refinements. For smaller sets (BAliBASE, PREFAB, SABmark) consistency introduced relevant information elevating result quality. Nevertheless, at the same time it interposed noise which accumulated for large sets of sequences causing accuracy decay even after the first iteration (800x2). Figure~\ref{Figure:consistency_details}c proves consistency to be harmful on large benchmarks independently of refinement iteration. Figure~\ref{Figure:consistency_scalability}a shows that the noise started to exceed positive signal for $k > 400$. 

Clearly, selecting only part of sequences for consistency can potentially increase the effectiveness of the procedure. To investigate a correlation between information content and evolutionary relationship of sequences involved, we applied triangle stochastic filters with an expected acceptance rate of 10\%. Those were low-pass, mid-pass and high-pass filters which promoted consistency over respectively, closely, mildly, and distantly related sequences. 
The shapes of the filter functions are presented in Supplementary Figure 2.
Distances were calculated as alignment scores from stage I ranked and normalised to $[ 0,1]$, the sum was used as $f$ function. Figure~\ref{Figure:consistency_details}c shows, that closely related sequences introduce more information to the consistency, thus should be preferred in the selection. Besides stochastic filtering, deterministic selectivity based on a structure of the guide tree was examined.
The consistency over sequence $z$ was performed when sum or maximum of tree-based distances $d_{xz}$ and $d_{yz}$ was smaller than assumed threshold $T$. 
The comparison of selectivity strategies (Figure~\ref{Figure:consistency_details}b) demonstrates deterministic variant with maximum function thresholded at $T \simeq 200$ to perform the best. It was superior to the version without consistency independently of refinement iteration with an exception of $r=0$ point where no-consistency won (Figure~\ref{Figure:consistency_details}d). The effect of consistency being profitable only when paired with refinement was not observed on smaller sets. To gain deeper insight into this phenomenon, more detailed investigation on interdependencies between consistency and refinement is required.

As a next step, we analysed the effect of amplification of the original $S_{xy}$ signal by multiplying its elements by coefficient  $h_{xy} \in [ 1, h ]$. The largest improvement in alignment quality was for $h = 3$ (see~Figure~\ref{Figure:consistency_details}d). 
This holds for all sets of 200 or more sequences (Figure~\ref{Figure:consistency_scalability}a) which coincides with the selectivity being configured to threshold maximum function at $T=200$ (expected number of relaxations for $k \geq 200$ is similar).
Results for BAliBASE, which contains much smaller sets, suggest that adjusting $h_{xy}$ individually for each $S_{xy}$ matrix depending to the number of performed relaxations works according to the expectations. Charts presenting influence of consistency on SP and TC measures separately can be found in Supplementary Figures~4 and~5.

The crucial feature of selective consistency is its computational scalability. For $k \geq 200$ an approximate number of relaxations for each posterior matrix is constant. As a result, time complexity of the procedure is $\Theta(k^2n^3)$ which is a noticeable improvement over full consistency variant. The comparison of execution times (Figure~\ref{Figure:consistency_scalability}b) shows that for large sets of sequences, time overhead related to selective consistency was negligible compared to other QuickProbs~2 stages.

\subsection*{Comparison with other algorithms}
The comparison of alignment software on benchmark datasets is given in Table~\ref{Table:comparison}. The algorithms were executed on \textit{desktop}\ configuration (see Table~\ref{Table:hardware} for details). Software packages suited for parallel processing were run with 12 processing threads to fully utilise multi-core architecture of the CPU.

\begin{table*}[t]
	\small
	\caption{Qualitative results for benchmark datasets on \textit{desktop} configuration. All versions of QuickProbs were run on Radeon 7970 due to incompatibility of QuickProbs 1 with GeForce 980. Best quality results typed in bold. Execution times given in $hh$:$mm$:$ss$ format. Dashed lines separates consistency (top) from (non-consistency) methods. 
		\label{Table:comparison}}%
	
	\renewcommand{\tabcolsep}{0.3em}
	\begin{tabular*}{\textwidth}{llrrrlrrrlrrrlrrrlrrrlrrr}
		\toprule
		Algorithm	&& \mc{3}{c}{BAliBASE}	&& \mc{3}{c}{PREFAB} 	&& \mc{3}{c}{OXBench-X} && \mc{3}{c}{SABmark}	&& \mc{3}{c}{BaliFam} 	&& \mc{3}{c}{HomFam$^\text{1K}$} \\
		\cline{3-5} 			\cline{7-9}				\cline{11-13}	\cline{15-17}	\cline{19-21}	\cline{23-25}
		&& time & SP & TC 		&& time & \mc{2}{c}{SP/TC} && time & SP & TC && time 	& SP & TC 		&&	time & SP & TC 		&& time & SP & TC  	\\
		\midrule
		QuickProbs~2	&& 2:01	& \tb{88.1} & \tb{61.8}	&& 8:18	&\mcc{\tb{74.2}} && 8:20 & \tb{89.4} & \tb{80.3}   && 10 & 61.1 & 40.7	&& 3:35:46 & \tb{84.7} & \tb{54.8}
		&&  1:43:36 & \tb{87.7} & \tb{72.0}\\ 
		QuickProbs-acc	&& 25:45 & 87.9 & 60.8	&& 57:25 &\mcc{74.0} 	&& 4:00:03 & 89.3 & 80.2 	&& 53 & 60.3 & 40.1			&& --- & --- & ---			&& --- & --- & --- \\
		QuickProbs		&& 5:17	 & 87.8 & 60.7	&& 15:37 &\mcc{73.6} 	&& 34:15	& 89.1 & 80.0 	&& 20 & 60.3 & 40.1			&& --- & --- & ---			&& --- & --- & --- \\
		MSAProbs		&& 25:12 & 87.8 & 60.8 	&& 1:42:51 &\mcc{73.7}	&& 1:55:04 	& 89.1 & 80.0 	&& 30 & 60.2 & 40.0 		&& $>$8\,days & 60.9 & 34.5 && 68:57:43 & 77.5 & 60.9\\
		PicXAA-PF		&& 3:20:51 & 87.8 & 59.3&& 13:31:09 &\mcc{71.2}	&& 17:09:16 & 88.3 & 78.4	&& 3:35	& 59.0 & 38.4		&& --- & --- & ---			&& --- & --- & --- \\
		PicXAA-HMM		&& 2:13:35 & 86.5 & 56.4&& 9:12:24 	&\mcc{71.1}	&& 13:17:57 & 87.8 & 77.4	&& 2:50	& 59.3 & 39.0		&& --- & --- & ---			&& --- & --- & --- \\
		GLProbs			&& 40:12   & 87.9 & 59.3&& 2:06:36  &\mcc{72.4}	&& 2:01:48	& 89.1 & 80.0  	&& 58 & \tb{61.4} & \tb{41.4}	&& --- & --- & ---		&& --- & --- & --- \\ 
		MAFFT-auto		&& 11:23   & 86.5 & 58.7&& 20:44 	&\mcc{72.6}	&& 7:56 	& 88.7 & 79.4	&& 1:28 & 57.3 & 36.8		&& 13:13 & 66.0 & 28.8		&& 14.15 & 81.5 & 62.1 \\ 
		\hdashline
		Clustal$\Omega$-iter2 && 4052 & 84.8 & 56.7	&& 9346	&\mcc{71.0}    	&&  2731    & 88.5 & 79.5	&& 172  & 55.2 & 35.7	&& 11:11:17 & 83.7 &	51.8&& 3:23:56 & 85.1 & 68.6\\
		Clustal$\Omega$	&& 4:56 	& 84.2 & 55.9	&& 14:19 &\mcc{70.0}	&& 4:45 	& 87.8 & 78.1 	&& 18	& 55.0 & 35.5	&& 1:27:21 & 79.9  & 44.5  	&& 46:10 & 84.1 & 67.2 \\
		MUSCLE			&& 8:47		& 81.9 & 47.8	&& 22:32 &\mcc{67.7}	&& 17:08 	& 87.5 & 77.6 	&& 32	& 54.5 & 33.5	&& $>$4\,days & 52.1 & 22.3	&& 31:42:23 & 70.6 & 50.8 \\
		Kalign-LCS		&& 21 		& 83.0 & 50.4	&& 1:30 &\mcc{65.9}		&& 27 		& 86.8 & 76.4 	&& 2	& 55.6 & 35.6	&& 6:36 & 67.5 & 31.5 		&& 4:14 & 81.5 & 62.1\\		
		MAFTT			&& 1:38		& 81.7 & 47.5	&& 8:03 &\mcc{68.0}		&& 2:10 	& 86.6 & 76.2 	&& 54	& 53.2 & 33.0	&& 13:13 & 66.0 &  28.8		&& 3:50 & 79.1 & 58.1\\
		Kalign2			&& 26 		& 81.1 & 47.1	&& 1:39 &\mcc{65.5}		&& 34 		& 86.3 & 75.6 	&& 2	& 52.4 & 32.6	&& 11:25 & 66.6 & 31.0  	&& 5:43 & 77.4 & 57.5\\		
		\bottomrule
	\end{tabular*}
\end{table*}

\begin{figure*}[t]
	\centering	
	\includegraphics[width=\textwidth]{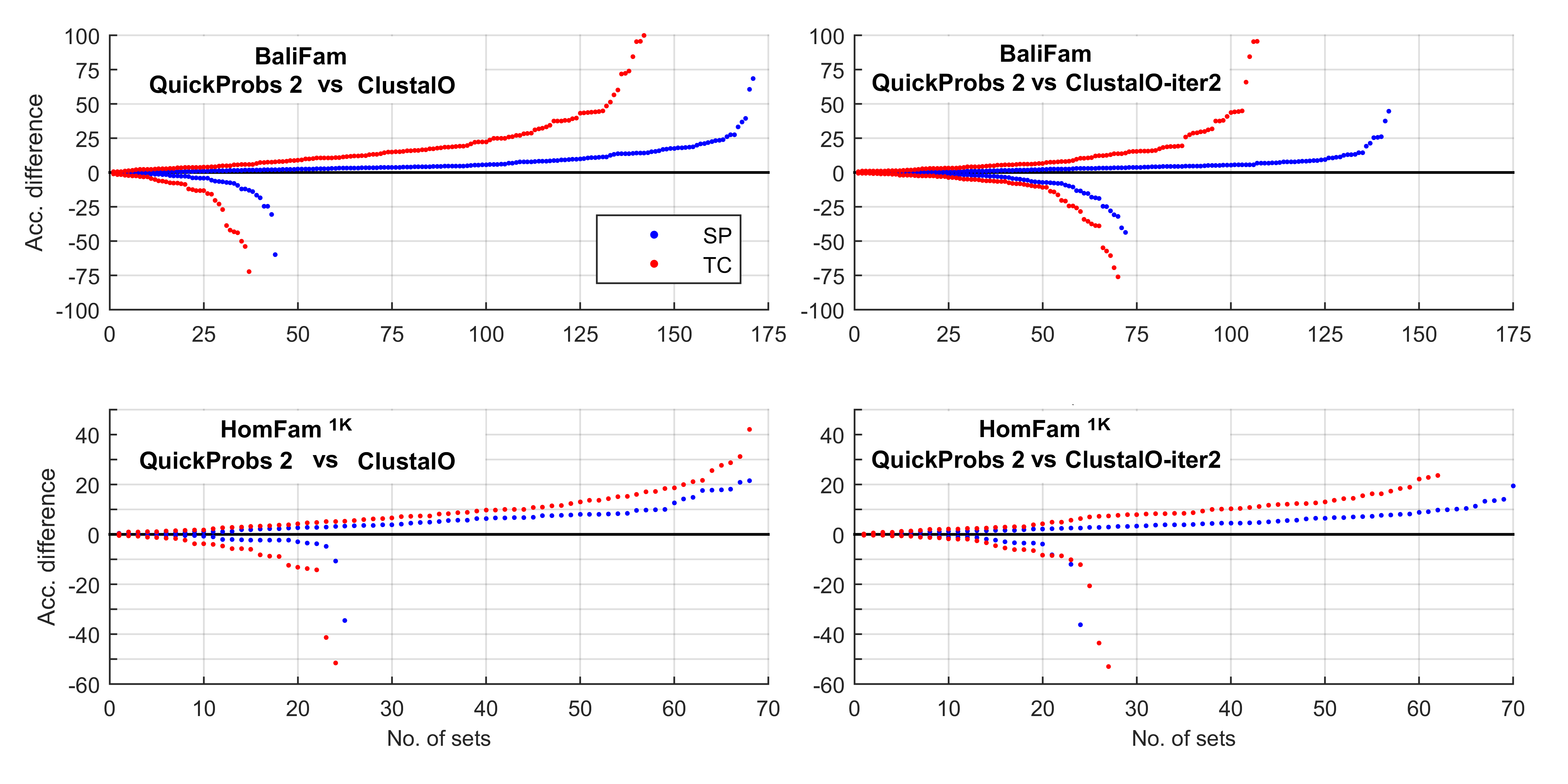}	
	\caption{Detailed comparison of QuickProbs~2 and Clustal$\Omega$ variants on BaliFam and HomFam$^\text{1K}$ benchmarks. For each quality measure (SP/TC) differences on individual protein families were sorted and plotted as two independent series. The points above the horizontal axis represent sets on which QuickProbs~2 was superior, the ones below correspond to the opposite situation.}
	\label{Figure:comparison_qp2}
\end{figure*}

For small sets of sequences (BAliBASE, PREFAB, OXBench, and SABmark) QuickProbs~2 competes with other consistency-based algorithms. Experiments show QuickProbs~2 to overcome them by a small margin (the distance to the second best does not exceed one percentage point on both SP and TC) with an exception of SABmark where GLProbs~\cite{Ye2015} took the lead. This can be explained by GLProbs being equipped in local alignment Markov models, which are especially profitable on distantly related sequences as those in SABmark. PicXAA, the only non-progressive algorithm in a comparison is also inferior to QuickProbs~2. In the case of large sets of sequences (BaliFam and HomFam$^\text{1K}$), consistency methods became inapplicable for real problems due to hardware limitations. Moreover, the accuracy of consistency deteriorated dramatically giving way to Clustal$\Omega$ and confirming the latter to be the good choice when numerous alignments are of interest. Nevertheless, thanks to column-oriented refinement and selective consistency, QuickProbs~2 was noticeably more accurate than the default mode of Clustal$\Omega$ on both large sets. E.g., the greatest advantage observed on BaliFam in TC score corresponds to almost 25\% more successfully aligned columns. 
When one considers Clustal$\Omega$ with two combined iterations enabled, QuickProbs~2 was still superior by a fair margin. Figure~\ref{Figure:comparison_qp2} presents a detailed comparison of the presented algorithm and Clustal$\Omega$ variants on BaliFam and HomFam$^\text{1K}$ benchmarks. For all families in a benchmark absolute advantages of QuickProbs over competing software in SP and TC measures were determined. For each measure, the differences were sorted and plotted on a chart as two independent series. The points above the horizontal axis represent sets on which QuickProbs~2 was superior, the ones below correspond to the opposite situation. This way one can asses on what portion of the dataset and to what extent one algorithm performed better than the other. The advance of QuickProbs~2 over default variant of Clustal$\Omega$ is clear: on both analysed benchmarks our algorithm was superior to the competitor on approximately 3/4 families. This was also the case for Clustal$\Omega$-iter2 on HomFam$^\text{1K}$. A bit different situation was for BaliFam, where enabling combined iterations noticeably improved Clustal$\Omega$ results. Though, it was still clearly inferior to QuickProbs~2.

\begin{figure*}[t]
	\centering	
	\includegraphics[width=0.5\textwidth]{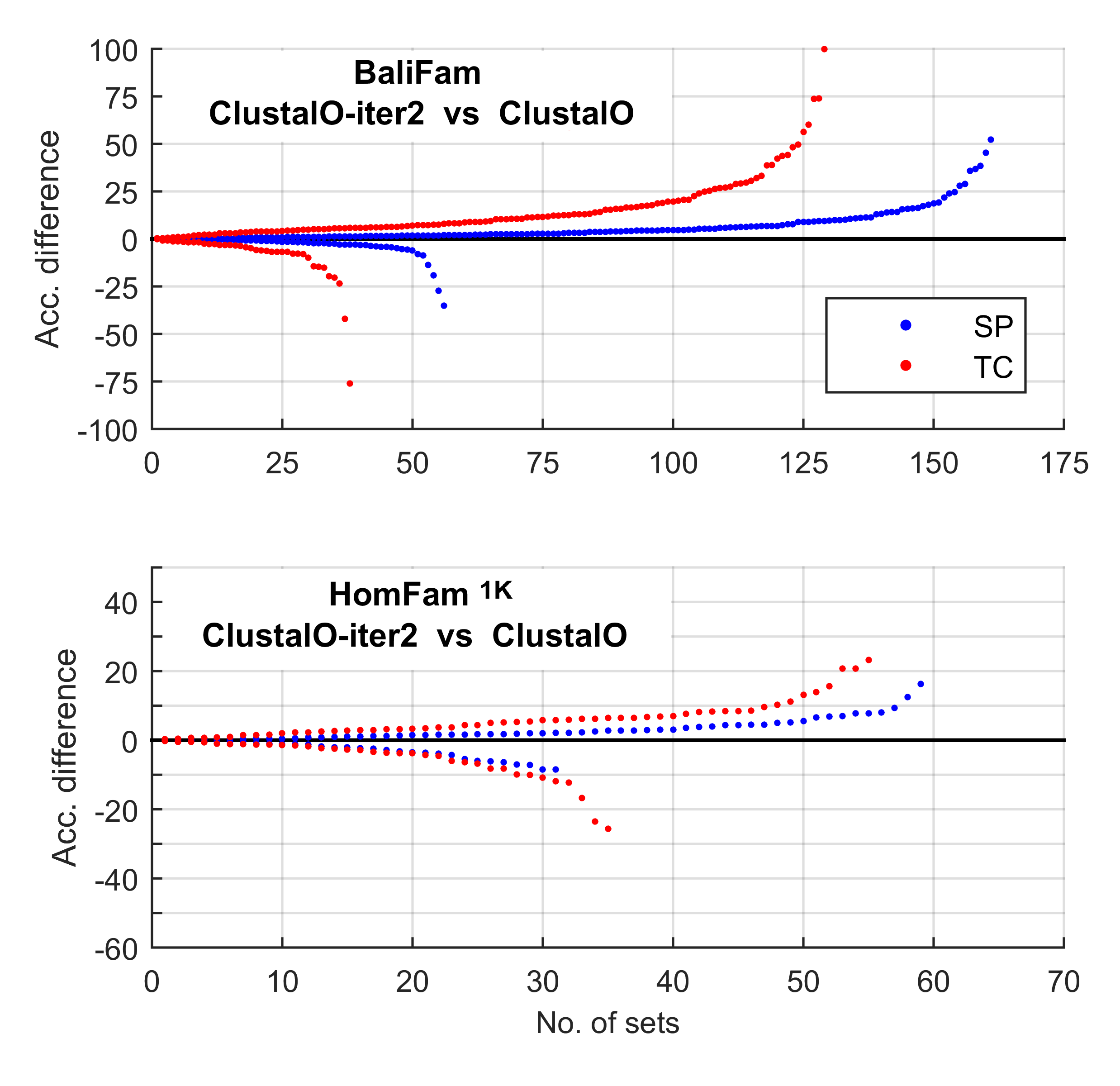}	
	\caption{Detailed comparison of Clustal$\Omega$-iter2 over Clustal$\Omega$ on BaliFam and HomFam$^\text{1K}$ benchmarks. For each quality measure (SP/TC) differences on individual protein families were sorted and plotted as two independent series. The points above the horizontal axis represent sets on which Clustal$\Omega$-iter2 was superior, the ones below correspond to the opposite situation.}
	\label{Figure:comparison_clustal}
\end{figure*}

The effect of presented algorithm being worse than Clustal$\Omega$ on several test cases is natural and is visible also when comparing other algorithms. For instance, combined iterations were reported to significantly elevate the quality of Clustal$\Omega$~\cite{Sievers2011, Sievers2013} results. However, when analyzing differences on particular protein families, there are sets for which default configuration is more accurate (Figure~\ref{Figure:comparison_clustal}). This is can be explained by the high diversity of alignment problems which hinders the development of algorithms superior to the competitors systematically on all test cases. Therefore, the statistical analysis of the results is necessary to properly assess performance of investigated methods. Significance of reported differences was verified with a use of Wilcoxon signed-rank test (Table~\ref{Table:significance}). To control family-wise error at $\alpha=0.05$, Bonferroni-Holm correction was applied. Low $p$-values for BaliFam and HomFam give strong evidence that QuickProbs~2 is currently the best algorithm for alignment of large sets of sequences also when compared to Clustal$\Omega$-iter2. The lack of significance was observed in few cases concerning small sets only (including the advantage of GLProbs over QuickProbs~2).

\begin{table}[t]
	\caption{Statistical significance of the results. $P$-values of Gm differences between QuickProbs~2 and selected methods were measured with a use of Wilcoxon signed-rank test. Insignificant results at $\alpha=0.05$ are given in parentheses (Bonferroni-Holm correction for multiple testing was applied), minus sign indicates the advantage of the competing software.}
	\label{Table:significance}
	\setlength{\tabcolsep}{0.0em}
	\begin{tabular*}{\hsize}{@{\extracolsep{\fill}}lllllllllll}
		\toprule
		Benchmark			&& QuickProbs-acc	&& MSAProbs 	&& GLProbs 			&& Clustal$\Omega$	&& Clustal$\Omega$-iter2\\
		\midrule
		BAliBASE			&& (0.02076)		&& (0.01352)	&& 0.00040			&& $<10^{-18}$	&& $<10^{-12}$\\
		PREFAB				&& (0.01694)		&& $<10^{-6}$ 	&& $<10^{-20}$		&& $<10^{-20}$	&& $<10^{-20}$\\
		OXBench-X			&& (0.00878)		&& 0.00169 		&& (0.29363)		&& $<10^{-9}$	&& 0.00538\\
		SABmark				&& 0.00425 			&& 0.00271 		&& ($-$0.04700)		&& $<10^{-14}$	&& $<10^{-13}$\\
		BaliFam				&& --- 				&& $<10^{-11}$ 	&& --- 				&& $<10^{-13}$	&& 0.00034\\		
		HomFam$^\text{1K}$	&& ---  			&& $<10^{-11}$	&& --- 				&& $<10^{-6}$	&& $<10^{-5}$\\	
		\bottomrule
	\end{tabular*}
\end{table}

\begin{table}[t]
\caption{Execution times of QuickProbs~2 in GPU and CPU modes. BaliFam and HomFam$^\text{1K}$ were not analysed on \textit{laptop} due to memory requirements. BAliBase, PREFAB, OXBench, and SABmark were processed in bulk mode.}
\label{Table:hardware}
\setlength{\tabcolsep}{0.0em}
\begin{tabular*}{\hsize}{@{\extracolsep{\fill}}llrrrlrrlrr}
	\toprule
	&& \mccc{\textit{desktop}}					&& \mcc{\textit{laptop}} 			&& \mcc{\textit{workstation}}\\
	\cline{3-5} 				\cline{7-8}				\cline{10-11}
	&& \mccc{i7 4930K (6$\times$3.4\,GHz)}		&& \mcc{i7-4700MQ (4$\times$2.4\,GHz)}		&& \mcc{2$\times$Xeon E5-2670v3 (24$\times$2.3\,GHz)}\\
	&& \mccc{64\,GB RAM}				&& \mcc{8\,GB RAM}				&&  \mcc{128\,GB RAM}\\
	&& \mccc{Radeon 7970$^{1}$ + GeForce 980$^{2}$}	&& \mcc{Radeon 8970M$^{3}$}		&&\mcc{Quadro M6000$^{4}$}\\
	\cline{3-5} 						\cline{7-8}							\cline{10-11}
	&& Radeon & GeForce  & CPU			&&  GPU & CPU		&&  GPU & CPU\\	
	\midrule
	BAliBASE		&&	 2:01 & 2:43 & 15:18			&& 2:55 & 23:04		&& 2:19 & 6:41\\  							
	PREFAB		&&  8:18 & 12:32 & 1:05:12 			&& 13:03 & 1:41:16	&& 9:15 & 26:43\\  
	OXBench		&&  8:20 & 11:37 & 57:20			&& 13:48 & 1:29:24	&& 10:25 & 25:15\\
	SABmark 			&&  10 & 18 & 27					&& 11 & 35			&& 18	& 20\\
	BaliFam	 	&&  3:35:46 & 5:15:40 & 34:25:29 	&& --- & ---		&& 5:13:22 & 14:00:03 \\
	HomFam$^\text{1K}$	&&  1:43:36 & 2:29:31 & 22:21:42	&& --- & ---		&& 2:29:05 & 6:32:54\\	
	\bottomrule
\end{tabular*}
\begin{itemize}[noitemsep,nolistsep]
	\item[$^{1}$]\,2048$\times$1.0\,GHz, 3\,GB RAM (228\,GB/s), 15.30 driver,  Windows 7\,x64
	\item[$^{2}$]\,2048$\times$1.2\,GHz, 4\,GB RAM (224\,GB/s), 359.06 driver, Windows 10\,x64
	\item[$^{3}$]\,1280$\times$0.9\,GHz, 4\,GB RAM (154\,GB/s), 15.11 driver,  Windows 7\,x64
	\item[$^{4}$]\,3072$\times$1.0\,GHz 12\,GB RAM (317\,GB/s), 352.55 driver, CentOS 7.1\,x64
\end{itemize}	
\end{table}
		
Superior accuracy of QuickProbs~2 on large protein families coincides with computational scalability. QuickProbs~2 is comparable to default mode of Clustal$\Omega$ in terms of execution times and orders of magnitude faster than consistency-based methods (MSAProbs needed over a week to complete BaliFam, QuickProbs 1 failed to run properly due to memory requirements). As QuickProbs~2 employs OpenCL, it can be executed on different massively parallel devices like NVidia and AMD GPUs. Moreover, presented software has also the ability to be run on central processor without OpenCL. As experiments on different hardware platforms show (Table~\ref{Table:hardware}), CPU variant is 3--10 times slower than GPU version, though still much faster than other algorithms based on consistency. 

\section*{Discussion}

Constantly growing availability of genomic and proteomic data opens new opportunities in life sciences. Yet, it is also a major challenge facing algorithms for sequence analyses, including multiple sequence alignment. Increasing number of sequences is one of the most important factors determining the difficulty of the MSA problem. In our research we have confirmed refinement and consistency, two most popular quality-aimed techniques employed by progressive aligners, to be ineffective or even harmful for sets of hundreds and more sequences. We present QuickProbs~2, a multiple alignment algorithm equipped with novel column-oriented refinement and selective consistency. It scales well with the number of sequences offering significantly better accuracy than Clustal$\Omega$---the previous leader for analysing large sets of sequences. For less numerous sets ($k < 100$), when methods based on full consistency like MSAProbs or PicXAA are applicable, QuickProbs~2 is still superior to the competitors. What is important, outstanding accuracy is obtained in a short time thanks to utilisation of massively parallel architectures.     

By successfully extending applicability of refinement and consistency to approximately thousand of sequences, we showed that sets of different sizes require various treatment. An open issue though, is the scalability of presented ideas for families of tens or hundreds thousands of sequences that are common in PFam database. This is caused by memory requirements of QuickProbs~2, the main issue to be resolved in future releases. For such large sets of sequences Clustal$\Omega$ or MAFFT are still the choice. 

Other factors contributing to the complexity of multiple alignment problem are sequence lengths, their evolutionary relationship, presence of long terminal fragments, etc. We believe that future development of MSA domain is impossible without better understanding of the influence of all these elements on alignment algorithms. Especially, in the light of recent discoveries made by Boyce et at.~\cite{Boyce2014} who showed that degenerated chained guide trees can elevate accuracy of selected alignment algorithms
when large sets of sequences are investigated. Our research also leads to some observations that remain to be explained, e.g., the effect of consistency being profitable for large protein families only when paired with refinement. Deeper involvement of biological community, which by definition is the major recipient of multiple alignment algorithms, would considerately facilitate advances in this area of computational biology.

QuickProbs~2 executables together with source code are available at \href{https://github.com/refresh-bio/QuickProbs}{https://github.com/refresh-bio/QuickProbs}
All examined datasets can be downloaded from \href{http://dx.doi.org/10.7910/DVN/7Z2I4X}{http://dx.doi.org/10.7910/DVN/7Z2I4X}. Web service for remote analyses is under development.

\bibliography{quickprobs-scirep}

\section*{Acknowledgements}
The work was supported by Polish National Science Centre upon decisions DEC-2012/05/N/ST6/03158 and DEC-2011/03/B/ST6/01588, Silesian University of Technology under BKM-507/RAU2/2016 project, performed using the infrastructure supported by POIG.02.03.01-24-099/13 grant: ``GeCONiI---Upper Silesian Center for Computational Science and Engineering''. We wish to thank Adam Adamarek for proofreading the manuscript.

\section*{Author contributions statement}
AG and SD designed the algorithm. AG implemented the algorithm. AG and SD designed and carried out the experiments. AG performed statistical analysis of the results. AD drafted the manuscript and supplementary material. Both authors read and approved the final manuscript.

\section*{Additional information}
The supplementary material contains information concerning employed benchmarks as well as extended experimental results.

\subsection*{Competing financial interests}
The authors declare no competing financial interests.


\end{document}